# Energy spectra and quadrupole transition probabilities of [124-130]Ba

H. Sabri[a*], M. Seidi[,b],

[a] Department of Physics, University of Tabriz, Tabriz 51664, Iran.
[b] Department of Physics, Ilam University, Ilam **516-69315**, Iran

---

[*] h-sabri@tabrizu.ac.ir

Abstract

In this paper, we have studied the energy spectra and $B(E2)$ transition probabilities of $^{124-130}$Ba isotopes in the shape phase transition region between the spherical and gamma unstable deformed shapes. We have used a transitional Interacting Boson Model Hamiltonian which is based on affine $SU(1,1)$ Lie Algebra in the both IBM-1 and 2 versions and also the Catastrophe theory in combination with a coherent state formalism to generate energy surfaces and determine the exact values of control parameters. Our results for control parameters suggest a combination of $U(5)$ and $SO(6)$ dynamical symmetries in this isotopic chain. Also, the theoretical predictions can be rather well reproduce the experimental counterparts when the control parameter is approached to the $SO(6)$ limit.

1. Introduction

The study of quantum phase transitions (QPTs) is a hot topic in different areas of quantum many-body physics. In Nuclear Physics many aspects of QPTs have been studied. When the numbers of protons (or neutrons) are modified, the energy levels and electromagnetic transition rates of atomic nuclei change too and suggest a transition from one kind of the collective behavior to another [1-3]. The quantum shape phase transitions have been studied 25 years ago with using the classical limits of the Interacting Boson Model (IBM) [4-10] which describes the nuclear structure of even–even nuclei within the $U(6)$ symmetry, possessing $U(5)$, $SU(3)$ and $O(6)$ dynamical symmetry limits, These descriptions point out that there is a first order shape phase transition between $U(5)$ and $SU(3)$ limits and a second order shape phase transition between $U(5)$ and $O(6)$ limits. The analytic description of nuclear structure at the critical point of phase transitions has attracted extensive interest in the recent decades. One has to employ some complicated numerical methods to diagonalize the transitional Hamiltonian in these situations but Pan *et al* in Refs.[11-12] have been proposed a new solution which was based on affine $SU(1,1)$ algebraic technique and explores the properties of nuclei have classified in the *U*(5)↔*SO*(6) transitional region of IBM.

It was long believed, the Barium isotopes were good examples of the quadrupole vibrational nuclei, namely *U*(5) nuclei [13-34]. However, during the last few years, new experimental data and calculations have led to a modified picture on these nuclei. By using the collective models in describing the structure of Barium isotopes [13], these nuclei can consider to be soft with regard to the $\gamma$ deformation with an almost maximum effective trixiality of $\gamma \approx 30°$. These mean *Ba* isotopes appear to evolve from the *U*(5) to *O*(6)-like structure in IBM classification. On the other hand, different authors who have been extensively studied the structure of *Ba* isotopes, consistently explained that the normal phonon states of this isotopic chain can be described via *U*(5) limit where one has to use the *O*(6) limit predictions for the intruder states. Also, the examinations of the three phonon states suggest that IBM-2 calculation were needed to explain these levels [15]. All these new experiments and theoretical calculations have provided new insights for these nuclei which is helpful to understand their structures [17-25]. Although, it is very difficult to treat them in terms of conventional mean field theories since they are neither vibrational nor rotational.

In this study, we have focused on the $^{124-130}Ba$ isotopes with emphasis on the energy levels and quadrupole transition probabilities. We have used the transitional Hamiltonian [35-36] to consider the evolution of these isotopes between spherical and gamma unstable shapes. Catastrophe formalism [37-44] is used to determine the exact values of control parameter for each nucleus and also present the energy surfaces. Different energy levels and quadrupole transition probabilities are determined in the IBM-1 and 2 frameworks and compared with experimental counterparts [45-49].

## 2. Theoretical framework

The phenomenological Interacting Boson Model (IBM) in terms of *U*(5), *SU*(3) and *O*(6) dynamical symmetries has been employed in describing the collective properties of several medium and heavy mass nuclei. These dynamical symmetries correspond to harmonic vibrator, axial rotor and $\gamma$ – unstable rotor as the geometrical analogues, respectively [1-3]. Although these symmetries are fairly successful in the investigation of low-lying nuclear states, the analytic description of structure at the critical point of phase transition is considered as issue, recently great analyses has been performed to describe them. Iachello in Refs.[1-2] have established a new set of dynamical symmetries, i.e. *E*(5) and *X*(5), for nuclei which are located at the critical point of transitional regions. The *E*(5) symmetry describes a second order phase transition which corresponds to the transitional states in the region from *U*(5) to *O*(6) symmetries in the IBM. Different analyses which have carried in the investigation of this transitional region [13-36] have suggested some nuclei such as $^{134}Ba$, $^{108}Pd$, as the empirical evidences for such a symmetry. Some complicated numerical calculation must be used to diagonalize the considered Hamiltonian in these transitional regions and critical points. To avoid these problems, an algebraic solution has been proposed by Pan *et al* [11-12] which was based on the affine $SU(1,1)$ Lie algebra to exhibits the properties of nuclei which are located in the *U*(5)↔*SO*(6) transitional region. Although the results of this approach are somewhat different from those of the IBM, but as have presented in Refs.[11-12,35-36], a clear correspondence with the description of the geometrical model is obvious for this transitional region.

2.1. Transitional Hamiltonian based on affine $SU(1,1)$ algebra

The $SU(1,1)$ Algebra has been described in detail in Refs.[11-12]. Here, we briefly outline the basic ansatz and summarize the results. The Lie algebra corresponds to the $SU(1,1)$ group is generated by $S^\nu$, $\nu = 0$ and $\pm$, which satisfies the following commutation relations

$$[S^0, S^\pm] = \pm S^\pm \quad , \quad [S^+, S^-] = -2S^0 \quad (1)$$

The Casimir operator of $SU(1,1)$ group can be written as

$$\hat{C}_2 = S^0(S^0 - 1) - S^+ S^- \quad , \quad (2)$$

Representations of $SU(1,1)$ are determined by a single number $\kappa$, thus the representation of Hilbert space is spanned by orthonormal basis $|\kappa\mu\rangle$ where $\kappa$ can be any positive number and $\mu = \kappa, \kappa+1,...$. Therefore,

$$\hat{C}_2(SU(1,1))|\kappa\mu\rangle = \kappa(\kappa-1)|\kappa\mu\rangle \qquad , \qquad S^0|\kappa\mu\rangle = \mu|\kappa\mu\rangle \tag{3}$$

In IBM, the generators of $d$ – boson pairing algebra is created by

$$S^+(d) = \frac{1}{2}(d^\dagger \cdot d^\dagger) \quad , \quad S^-(d) = \frac{1}{2}(\tilde{d} \cdot \tilde{d}) \quad , \quad S^0(d) = \frac{1}{4}\sum_\nu (d_\nu^\dagger d_\nu + d_\nu d_\nu^\dagger) \tag{4}$$

Similarly, $s$ – boson pairing algebra forms another $SU^s(1,1)$ algebra which is generated by

$$S^+(s) = \frac{1}{2}s^{\dagger 2} \quad , \quad S^-(s) = \frac{1}{2}s^2 \quad , \quad S^0(s) = \frac{1}{4}(s^\dagger s + ss^\dagger) \tag{5}$$

On the other hand, the infinite dimensional $SU(1,1)$ algebra is generated by using of [11-12]

$$S_n^\pm = c_s^{2n+1} S^\pm(s) + c_d^{2n+1} S^\pm(d) \qquad , \qquad S_n^0 = c_s^{2n} S^0(s) + c_d^{2n} S^0(d) \tag{6}$$

Where $c_s$ and $c_d$ are real parameters and $n$ can be $0, \pm 1, \pm 2, \ldots$. These generators satisfy the commutation relations,

$$[S_m^0, S_n^\pm] = \pm S_{m+n}^\pm \qquad , \qquad [S_m^+, S_n^-] = -2S_{m+n+1}^0 \tag{7}$$

Then, $\{S_m^\mu, \mu = 0, +, -; \pm 1, \pm 2, \ldots\}$ generates an affine Lie algebra $SU(1,1)$ without central extension. By employing the generators of $SU(1,1)$ Algebra, the following Hamiltonian is constructed for the transitional region between $U(5) \leftrightarrow SO(6)$ limits [11-12]

$$\hat{H} = g\, S_0^+ S_0^- + \varepsilon\, S_1^0 + \gamma\, \hat{C}_2(SO(5)) + \delta\, \hat{C}_2(SO(3)) \tag{8}$$

$g, \varepsilon, \gamma$ and $\delta$ are real parameters where $\hat{C}_2(SO(3))$ and $\hat{C}_2(SO(5))$ denote the Casimir operators of these groups. It can be seen that Hamiltonian (8) would be equivalent with $SO(6)$ Hamiltonian if $c_s = c_d$ and with $U(5)$ Hamiltonian when $c_s = 0\ \&\ c_d \neq 0$. Therefore, the $c_s \neq c_d \neq 0$ requirement just corresponds to the $U(5) \leftrightarrow SO(6)$ transitional region. In our calculation we take $c_d (=1)$ constant value and $c_s$ vary between 0 and $c_d$.

Eigenstates of Hamiltonian (8) can obtain with using the Fourier-Laurent expansion of eigenstates and $SU(1,1)$ generators in terms of unknown $c$ – number parameters $x_i$ with $i = 1, 2, \ldots, k$. It means, one can consider the eigenstates as [11-12]

$$|k;v_s v n_\Delta LM\rangle = \sum_{n_i \in Z} a_{n_1} a_{n_2}...a_{n_k} x_1^{n_1} x_2^{n_2}...x_k^{n_k} S_{n_1}^+ S_{n_2}^+...S_{n_k}^+ |lw\rangle \qquad , \qquad (9)$$

Due to the analytical behavior of wavefunctions, it suffices to consider $x_i$ near zero. With using the commutation relations between the generators of $SU(1,1)$ Algebra, i.e. Eq.(7), wavefunctions can be considered as:

$$|k;v_s v n_\Delta LM\rangle = N S_{x_1}^+ S_{x_2}^+...S_{x_k}^+ |lw\rangle \qquad , \qquad (10)$$

where $N$ is the normalization factor and

$$S_{x_i}^+ = \frac{c_s}{1-c_s^2 x_i} S^+(s) + \frac{c_d}{1-c_d^2 x_i} S^+(d) \qquad , \qquad (11)$$

The c-numbers $x_i$ are determined through the following set of equations

$$\frac{\epsilon}{x_i} = \frac{gc_s^2(v_s + \frac{1}{2})}{1-c_s^2 x_i} + \frac{gc_d^2(v + \frac{5}{2})}{1-c_d^2 x_i} - \sum_{i \neq j} \frac{2}{x_i - x_j} \qquad \text{for i=1,2,...,k} \qquad (12)$$

Eigenvalues of Hamiltonian (8), i.e. $E^{(k)}$, can be expressed as [11-12]

$$E^{(k)} = h^{(k)} + \gamma v(v+3) + \delta L(L+1) + \varepsilon \Lambda_1^0 \qquad , \qquad \Lambda_1^0 = \frac{1}{2}[c_s^2(v_s + \frac{1}{2}) + c_d^2(v + \frac{5}{2})] \qquad (13)$$

Which

$$h^{(k)} = \sum_{i=1}^{k} \frac{\varepsilon}{x_i} \qquad , \qquad (14)$$

The quantum number $k$, is related to total boson number $N$, by

$$N = 2k + v_s + v$$

To obtain the numerical results for $E^{(k)}$, we have followed the prescriptions have introduced in Refs.[7-8], namely a set of non-linear Bethe-Ansatz equations (BAE) with $k$ – unknowns for $k$ – pair excitations must be solved. To this aim we have changed the variables as

$$\epsilon = \frac{\varepsilon}{g}(g=1 \ kev \ [11\text{-}12]) \qquad c = \frac{c_s}{c_d} \leq 1 \qquad y_i = c_d^2 x_i$$

so, the new form of Eq.(12) would be

$$\frac{\epsilon}{y_i} = \frac{c^2(\nu_s + \frac{1}{2})}{1-c^2 y_i} + \frac{(\nu + \frac{5}{2})}{1-y_i} - \sum_{i \neq j} \frac{2}{y_i - y_j} \qquad \text{for i=1,2,...,k} \qquad (15)$$

We have solved Eq. (15) with definite values of $c$ and $\varepsilon$ for $i=1$ to determine the roots of Beth-Ansatz equations (BAE) with specified values of $\nu_s$ and $\nu$, similar to procedure which have done in Refs.[7-8]. Then, we have used "Find root" in the Maple13 to get all $y_j^{'}$s. We carry out this procedure with different values of $c$ and $\varepsilon$ to provide energy spectra (after inserting $\gamma$ and $\delta$) with minimum variation as compared to the experimental counterparts;

$$\sigma = (\frac{1}{N_{tot}} \sum_{i, tot} |E_{\exp}(i) - E_{cal}(i)|^2)^{1/2}$$

which $N_{tot}$ is the number of energy levels where are included in extraction processes. We have extracted the best set of Hamiltonian's parameters, i.e. $\gamma$ and $\delta$, via the available experimental data [27-29] for excitation energies of selected states, $0_1^+, 2_1^+, 4_1^+, 0_2^+, 2_2^+, 4_2^+$ and *etc*, e.g. 12 levels up to $2_4^+$, or two neutron separation energies for nuclei which are considered in this study. In summary, we have extracted $\gamma$ and $\delta$ externally from empirical evidences and other quantities of Hamiltonian, e.g. $c$ and $\varepsilon$ would determine through the minimization of $\sigma$.

As have been explained in Refs.[23-29], one may use the IBM-2 calculation to explore three phonons or 2p-4h states, e.g. intruder states. To this aim we have used same formalism to extend IBM-2 calculation via $SU(1,1)$ lie algebra. Details have been presented in Ref.[11-12] and we explore the final results. In IBM-2 case, the Hamiltonian can be considered as

$$\hat{H} = g\, S_0^+ S_0^- + \varepsilon\, S_1^0 + \gamma_1\, \hat{C}_2(SO_\pi(5)) + \gamma_2\, \hat{C}_2(SO_\nu(5)) + \delta_1\, \hat{C}_2(SO_\pi(3)) + \delta_2\, \hat{C}_2(SO_\nu(3)) + \delta\, \hat{C}_2(SO(3)) \qquad (16)$$

In this Hamiltonian, we have

$$S_n^\pm = \sum_t c_{s;t}^{2n+1} S^\pm(s;t) + c_{d;t}^{2n+1} S^\pm(d;t) \qquad , \qquad S_n^0 = \sum_t c_{s;t}^{2n} S^0(s;t) + c_{d;t}^{2n} S^0(d;t) \qquad (17)$$

And the sum is over proton, $\pi$, and neutron, $\nu$, indices. The eigenstates of Eq. (16) can be expressed as

$$|k;\beta;v_s^\pi,v_s^\nu,v^\pi,v^\nu;n_\Delta^\pi L_\pi,n_\Delta^\nu L_\nu;LM\rangle = NS_{x_1}^+ S_{x_2}^+ ... S_{x_k}^+ |lw\rangle \qquad (18)$$

where $2k = N_\pi + N_\nu - v_s^\pi - v_s^\nu - v^\pi - v^\nu$ and

$$S_{x_i}^+ = \sum_t \frac{c_{s;t}}{1-c_{s;t}^2 x_i} S^+(s;t) + \frac{c_{d;t}}{1-c_{d;t}^2 x_i} S^+(d;t) \qquad (19)$$

Similar to IBM-1 case, the $c$-number $x_i$ satisfy a set of equations similar to (12)

$$\frac{\epsilon}{x_i} = \sum_t g\left(\frac{c_{s;t}^2(v_s^t + \frac{1}{2})}{1-c_{s;t}^2 x_i} + \frac{c_{d;t}^2(v^t + \frac{5}{2})}{1-c_{d;t}^2 x_i}\right) - \sum_{i\neq j}\frac{2}{x_i - x_j} \qquad \text{for i=1,2,...,k} \qquad (20)$$

Finally, the eigenvalues of Eq.(16) can be expressed as

$$E^{(k)} = \sum_{i=1}^{k}\frac{\varepsilon}{x_i} + \gamma_1 v^\pi(v^\pi+3) + \gamma_2 v^\nu(v^\nu+3) + \delta_1 L_\pi(L_\pi+1) + \delta_2 L_\nu(L_\nu+1) + \delta L(L+1) + \varepsilon \Lambda_1^0$$

$$\Lambda_1^0 = \sum_t \frac{1}{2}[c_{s;t}^2(v_s^t + \frac{1}{2}) + c_{d;t}^2(v^t + \frac{5}{2})] \qquad (21)$$

Similar to procedures which have done to extract parameters of transitional Hamiltonian in IBM-1 framework, we have supposed $c_d = 1$ and then, Eq. (20) have solved for $i=1$ case with definite values of $c$ and $\varepsilon$. Other parameters of Hamiltonian, namely $\delta$ and $\gamma$, have extracted from empirical available data for isotopic chain and we would repeat these processes with different values of considered quantities to obtain the smallest $\sigma$ values.

## 2.2. Energy surfaces

We can investigate the geometric configuration of the considered model in the framework of coherent state. The coherent state formalism of IBM [37-44] connects the algebraic and geometric descriptions of three dynamical symmetry limits and also allows the study of transitions among them. by using this formalism, one can evaluate the ground state energy as a function of shape variables $\beta$ and $\gamma$, i.e. deformation parameters[37], similar to what have been done for $U(5) \leftrightarrow SO(6)$ and $U(5) \leftrightarrow SU(3)$ phase transitions [35,36]. The classical limit corresponding to Hamiltonian (2.8) is obtained by considering its expectation value in the coherent state [37-41]

$$|N,\alpha_m\rangle = (s^\dagger + \sum_m \alpha_m d_m^\dagger)^N |0\rangle \quad , \tag{22}$$

Where $|0\rangle$ is the boson vacuum state, $s^\dagger$ and $d^\dagger$ are the boson operators of the IBM, and parameter $\alpha_m$ can be related to the deformation collective parameters [37],

$$\alpha_0 = \beta \cos\gamma \quad , \quad \alpha_{\pm 1} = 0 \quad , \quad \alpha_{\pm 2} = \frac{\beta}{\sqrt{2}} \cos\gamma \tag{23}$$

In the IBM-2 framework, the most general form of coherent state is [42-44]

$$|N_\pi, N_\nu, \beta_\pi, \gamma_\pi, \beta_\nu, \gamma_\nu, \varphi, \theta, \psi\rangle = \frac{1}{\sqrt{(N_\pi)!(N_\nu)!}} R(\theta,\varphi,\psi)(\Gamma_\pi^\dagger)^{N_\pi} (\Gamma_\nu^\dagger)^{N_\nu} |0\rangle \quad , \tag{24}$$

where

$$\Gamma_\rho^\dagger = \frac{[s_\rho^\dagger + \beta_\rho \cos\gamma_\rho d_{\rho,0}^\dagger + \frac{1}{\sqrt{2}}\beta_\rho \sin\gamma_\rho (d_{\rho,2}^\dagger + d_{\rho,-2}^\dagger)]}{\sqrt{1+\beta_\rho^2}} \quad , \tag{25}$$

And the Euler angles, $(\theta,\varphi,\psi)$, define the orientation of deformation variables, $(\beta_\pi, \gamma_\pi)$ for proton bosons and $(\beta_\nu, \gamma_\nu)$ for neutron bosons which as has shown in Ref,[42], in the absence of hexadecupole interactions, one can take the Euler angels equal to zero. Energy surface would determine by means of

$$E = \frac{\langle N,\alpha_m | H | N,\alpha_m \rangle}{\langle N,\alpha_m | N,\alpha_m \rangle} \quad , \tag{26}$$

Then, the energy surfaces from each part of transitional Hamiltonian can be written as

$$<gS_0^+ S_0^-> = \frac{g}{4}\left(\frac{N_\rho(N_\rho-1)}{(1+\beta_\rho^2)^2}\right)(c_s^2 + 2c_s c_d \beta_\rho^2 + c_d^2 \beta_\rho^4) \quad , \tag{27}$$

$$<\alpha S_0^1> = \frac{\alpha c_s^2}{4}\left(\frac{2N_\rho}{1+\beta_\rho^2}+1\right) + \frac{\alpha c_d^2}{4}\left(\frac{2N\beta_\rho^2}{1+\beta_\rho^2}+5\right) \quad , \tag{28}$$

$$<\gamma \hat{C}_2(SO_\rho(5))> = 2\frac{\gamma N_\rho \beta_\rho^2}{1+\beta_\rho^2} \quad , \tag{29}$$

$$<\delta \hat{C}_2(SO_\rho(3))> = \frac{3}{5}\frac{\delta_\rho N_\rho \beta_\rho^2}{1+\beta_\rho^2} \quad , \tag{30}$$

These yields the energy surfaces in the IBM-1 framework as

$$E(\beta,\gamma) = \frac{g}{4}\left(\frac{N(N-1)}{(1+\beta^2)^2}\right)(c_s^2 + 2c_s c_d \beta^2 + c_d^2 \beta^4) + \frac{\alpha c_s^2}{4}\left(\frac{2N}{1+\beta^2}+1\right) + \frac{\alpha c_d^2}{4}\left(\frac{2N\beta^2}{1+\beta^2}+5\right) +$$
$$+2\frac{\gamma N\beta^2}{1+\beta^2} + \frac{3}{5}\frac{\delta N\beta^2}{1+\beta^2} \quad , \tag{31}$$

And similarly, we can get the energy surfaces in the IBM-2 framework

$$E(\beta,\gamma) = \frac{g}{4}(\frac{N_\rho(N_\rho-1)}{(1+\beta_\rho^2)^2})(c_s^2 + 2c_s c_d \beta_\rho^2 + c_d^2 \beta_\rho^4) + \frac{\alpha c_s^2}{4}(\frac{2N_\rho}{1+\beta_\rho^2}+1) + \frac{\alpha c_d^2}{4}(\frac{2N_\rho \beta_\rho^2}{1+\beta_\rho^2}+5) +$$

$$+2\frac{\gamma_1 N_\pi \beta_\pi^2}{1+\beta_\pi^2} + 2\frac{\gamma_2 N_\nu \beta_\nu^2}{1+\beta_\nu^2} + \frac{3}{5}\frac{\delta_1 N_\pi \beta_\pi^2}{1+\beta_\pi^2} + \frac{3}{5}\frac{\delta_2 N_\nu \beta_\nu^2}{1+\beta_\nu^2} + \frac{3}{5}\frac{\delta N_\rho \beta_\rho^2}{1+\beta_\rho^2} \qquad , \qquad (32)$$

To analyze the energy surfaces within the catastrophe theory formalism, which are explained completely in the Refs.[37-44], we have determined the critical points of the energy surfaces. The following algebraic equation is yield the variable $\beta$ ( in the IBM-1 formalism which the procedure is similar for IBM-2 and we will denote the final result)

$$\frac{\partial E}{\partial \beta} = \frac{\beta}{(1+\beta^2)^3}[gN(N-1)(c_s+c_d)(c_d-c_s)\beta^2 + 2(\frac{N}{2}\alpha c_d^2 + 2\gamma N + \frac{3}{5}\delta N - \frac{N}{2}\alpha c_s^2)(1+\beta^2)] \quad , \qquad (33)$$

Which will use to obtain the critical points. This expression show, the $\beta = 0$ is a critical point for any values of the parameters of the energy surfaces and is the fundamental root. The Taylor series expansion of the energy surfaces around this fundamental root is given by

$$E(\beta) = \frac{g}{4}N(N-1)c_s^2 + \frac{N}{2}\alpha c_s^2 + \frac{1}{4}\alpha(c_s^2 + 5c_d^2) +$$
$$+\frac{1}{2}[N(N-1)gc_s(c_d-c_s) + N(\alpha(c_d^2 - c_s^2) + \frac{6}{5}\delta + 4\gamma)]\beta^2 +$$
$$+[\frac{3}{4}N(N-1)gc_s^2 - N(N-1)gc_s c_d + \frac{1}{4}N(N-1)gc_d^2 + \frac{1}{2}N\alpha(c_s^2 - c_d^2) - \frac{3}{5}N\delta - 2N\gamma]\beta^4$$
$$+O(5) + ... \quad , \qquad (34)$$

Or can be rewritten in the form

$$E(\beta) = A + A'\beta^2 + A''\beta^4 + ... \qquad , \qquad (35)$$

While the coefficients are given by

$$A = \frac{g}{4}N(N-1)c_s^2 + \frac{N}{2}\alpha c_s^2 + \frac{1}{4}\alpha(c_s^2 + 5c_d^2) \qquad , \qquad (36)$$

$$A' = \frac{1}{2}[N(N-1)gc_s(c_d-c_s) + N(\alpha(c_d^2 - c_s^2) + \frac{6}{5}\delta + 4\gamma) \qquad \text{For IBM-1} \qquad (37)$$

$$A' = \frac{1}{2}[N(N-1)gc_s(c_d-c_s) + N(\alpha(c_d^2 - c_s^2) + \frac{6}{5}(\delta_1 + \delta_2) + 4(\gamma_1 + \gamma_2 + \gamma)) \qquad \text{For IBM-2} \qquad (38)$$

$$A^{"} = \frac{3}{4}N(N-1)gc_s^2 - N(N-1)gc_sc_d + \frac{1}{4}N(N-1)gc_d^2 + \frac{1}{2}N\alpha(c_s^2 - c_d^2) - \frac{3}{5}N\delta - 2N\gamma) \quad \text{For IBM-1} \quad (39)$$

$$A^{"} = \frac{3}{4}N(N-1)gc_s^2 - N(N-1)gc_sc_d + \frac{1}{4}N(N-1)gc_d^2 + \frac{1}{2}N\alpha(c_s^2 - c_d^2) -$$

$$-\frac{3}{5}(N_\pi\delta_1 + N_\nu\delta_2) - 2(N\gamma + N_\pi\gamma_1 + N_\nu\gamma_2)) \qquad \text{For IBM-2} \qquad (40)$$

We must determine the Bifurcation set, the locus of the points in the space of control parameters at which a transition occurs from one local minimum to another[8], to identify the exact value of control parameter for each nucleus. With using the *det(H)=0* condition, H is the matrix of the second derivate of the energy surface at the critical point, which became as $\partial^2 E/\partial\beta^2 = 0$ in the case of a function of one variable [39]. one gets the expression

$$c_s = \frac{g(N-1)c_d + \sqrt{g^2(N-1)^2 c_d^2 + 4[g(N-1)+\alpha][\alpha c_d^2 + 4\gamma + \frac{6}{5}\delta]}}{2(g(N-1)+\alpha)}, \qquad (41)$$

In the IBM-1 formalism and similarly we get the following result in the IBM-s formalism

$$c_s = \frac{g(N-1)c_d + \sqrt{g^2(N-1)^2 c_d^2 + 4[g(N-1)+\alpha][\alpha c_d^2 + 4(\gamma_1 + \gamma_2) + \frac{6}{5}(\delta_1 + \delta_2 + \delta)]}}{2(g(N-1)+\alpha)}. \qquad (42)$$

## 2.2. $B(E2)$ Transition

The reduced electric quadrupole transition probabilities, $B(E2)$, are considered as the observables which as well as quadrupole moment ratios within the low-lying state bands prepare more information about the nuclear structure. The *E2* transition operator must be a Hermitian tensor of rank two and consequently, number of bosons must be conserved. With these constraints, there are two operators possible in the lowest order, therefore the electric quadrupole transition operator employed in this study is defined as [7],

$$\hat{T}_\mu^{(E2)} = q_2 [\hat{d}^\dagger \times \tilde{s} + \hat{s}^\dagger \times \tilde{d}]_\mu^{(2)} + q_2' [\hat{d}^\dagger \times \tilde{d}]_\mu^{(2)}, \qquad (43)$$

Where $q_2$ is the effective quadrupole charge, $q_2'$ is a dimensionless coefficient and $s^\dagger(d^\dagger)$ represent the creation operator of $s(d)$ boson. Reduced electric quadrupole transition rate between $I_i \to I_f$ states is given by [3]

$$B(E2; I_i \to I_f) = \frac{\left|\langle I_f \| T(E2) \| I_i \rangle\right|^2}{2I_i + 1} \quad , \tag{44}$$

To analyze the *B(E2)* transition ratios for isotopic chain, we have calculated the matrix elements of *T(E2)* operator between considered states, then with comparing the results with the experimental counterparts, we can extract $(q_2, q_2')$ quantities. To this aim and also to simplify the description, we have followed the method introduced in Refs.[7-8] and in the fitting procedures, these parameters would be described as a function of only, total boson number $(N)$. On the other hand, In IBM-2 framework, *B(E2)* transition probability were calculated via *E2* operator

$$T(E2) = e_\pi Q_\pi + e_\nu Q_\nu \quad , \quad Q_\rho = q_2 [d_\rho^\dagger s_\rho + s_\rho^\dagger \tilde{d}_\rho]^{(2)} + q_2' [d_\rho^\dagger \tilde{d}_\rho]^{(2)} \quad \rho = \pi \,\&\, \nu \tag{45}$$

$e_\pi$ and $e_\nu$ are the effective charges for the proton and neutron bosons, respectively. These quantities were determined through a fitting procedure that includes all known levels and selected *B(E2)* transition probabilities. In our calculation and then comparison with the experimental counterparts, we explore only $B(E2; 4_1^+ \to 2_1^+)$ and $B(E2; 2_1^+ \to 0_1^+)$ transitions (and their ratios in different isotopes) and therefore intruder states would not have any effects in our results. On the other hand, for other transition ratios such as $B(E2; 0_2^+ \to 2_1^+)$ and $B(E2; 2_2^+ \to 0_1^+)$ which includes transition between some of intruder states, our model has some unusual variation in comparison with experimental counterparts and therefore, we would not consider them.

## 3. Numerical result

### 3.1. Energy levels

Investigations of experimental energy spectra which have been done in Refs.[13-34], suggest $^{124-130}Ba$ isotopes as the empirical evidences for *U(5)↔SO(6)* transitional region. Consequently, the transitional Hamiltonians, Eq.(8) in IBM-1 framework and Eq.(16) for IBM-2 calculations, can be considered in the determination of energy spectra. There are 12 levels up to the $2_4^+$ level for each nucleus which are included in the extraction procedure as displayed in Figures 1 and 2 for considered isotopic chain, respectively for IBM-1 and 2 predictions. The best fits for IBM-1 Hamiltonian's parameters, namely $\varepsilon, \delta$ and $\gamma$ which are extracted from experimental data, by similar method has been explained in Refs.[15-18], and then by using these quantities in Eq.39, we have determined the $c_s$ values which all of them are presented in Table 1. These quantities described the best agreement between the calculated energy levels in this study and their experimental counterparts taken from Refs.[42-46], i.e. minimum values for $\sigma$.

Figure1 describes a comparison between the available experimental levels and the predictions of our results for $^{124-130}Ba$ isotopes in the low-lying region of spectra. An acceptable degree of agreement is obvious between them.

Our results which suggest a combination of the vibrational and gamma unstable limits in Barium isotopes, e.g. $c_s$ values are in the middle range of allowed values and do not approach to one limit explicitly, confirm previous result which suggests a shape coexistence in the Barium isotopic chain. Also these results for $c_s$ verify the transitional behavior for these nuclei and therefore, suggest the application of these transitional Hamiltonian for such nuclei which are located in the near of closed shell. These results, similar to results which we have obtained in Ref.[36] for $Te$ isotopes, suggest a deviation from the $U(5)$ limit ($c_s = 0$) in this isotopic changes where our results for $c_s$ are located in the 0.38 - 0.60 region. Also, when the deformation effect is increased in our calculation ($c_s$ approaches to $SO(6)$ limit), the accuracy of our determination also increased where the smallest uncertainty is reported for $^{130}Ba$ which $c_s = 0.6$ is offered for it.

The existence of intruder states in some nuclei such as $Cd$, $Te$ isotopes which are located in the near of closed shells are reported in different studies [21-29]. This means, the normal vibrational construction would not explore the observed data for two phonon triplet states, $2p - 4h$ excitation. Different authors suggested to use the IBM-2 predictions to calculate separately the normal and intruder states. To this aim, we have introduced IBM-2 formalism of transitional Hamiltonian in previous sections. With using same procedure which has done for IBM-1 calculations to extract parameters of transitional Hamiltonian through experimental counterparts via Bethe-Ansatz method and Least Square Fitting (LSF) extraction, we have got the following results for these parameters which are presented in Table 2. Also a comparison between IBM-2 predictions and experimental counterparts for $^{124-130}Ba$ are given in Figure 2.

IBM-2 suggests more exact results, i.e. minimum $\sigma$ values, in comparison with experimental data and therefore, one can consider this framework to describe energy spectra of *Ba* isotopes. Figures 1 and 2 explore the ability of $SU(1,1)$ based transitional Hamiltonian in both IBM-1 and 2 formalisms in reproduction of all considered levels in this study and also the acceptable degree of extraction procedures.

The results of IBM-2 calculation, confirm our idea about the effect of deformation in results to reduce the distance between theoretical predictions and experimental counterparts. In comparison with predictions of IBM-1, we reach to so low uncertainty but our results for $c_s$ values are more than IBM-1 counterparts. Also, similar to results of IBM-1, we observe the minimum $\sigma$ values for nuclei which our results suggest the maximum $c_s$ values for them.

In this paper, we have considered the energy surfaces for *Ba* isotopes, too. To obtain them, we have used the results which are yield via IBM-1 predictions. In the transition from $U(5)$ to the $SO(6)$ limit, the evolution of energy surface goes from a pure $\beta^2$ to a combination of $\beta^2$ and $\beta^4$ that has a deformed minimum [37]. Figure 3 shows the energy surfaces for the *Ba* isotopic chain which are plotted as a function of $\beta$. Our results in Table 2 propose the most $SO(6)$ - like structure for $^{126}Ba$ nucleus and if we consider the energy surfaces for different dynamical symmetry limits which are presented in Ref.[37], this isotope has the most like shape to the expected energy surface for $SO(6)$ limit. We do not represent the same comparison for energy surfaces by IBM-2 predictions where the variation of both $\beta_\pi$ and $\beta_\nu$ produce some 3D figures which any remarkable result can not realize from them.

In our considered framework, we have compared the predictions of transitional Hamiltonian, in both IBM-1 and IBM-2 cases, for energy spectra with their experimental counterparts, which contain all the observed intruder states [13-34]. Also, we have tried to extract the best set of parameters which reproduce these complete spectra with minimum variations. It means, our suggestion to use this transitional Hamiltonian for description of *Ba* isotopic chain would not has any contradiction with other theoretical studies which have been done with special hypotheses about mixing of intruder and normal configurations.

Our model has proposed the $c_s$ values in the 0-1 region, without any dominant deviation to any limits, and confirms the mixing of both vibrating and rotating structures in these nuclei. This may suggest the same role of mixing parameter of other investigation which explain the combination of normal and intruder configurations to our control parameter. On the other hand, to appreciate the advantages (or disadvantages) of different theoretical methods, we have compared our results with different analyses such as Refs.[14-25]. Our selected nuclei have not considered in other theoretical studies completely, but for some of them which we have found similar counterparts, our results which have derived via more parameters (as compared to standard IBM shape-phase transition approaches) explore the smallest $\sigma$ values. This means, for these numbers of levels in this energy region, the affine $SU(1,1)$ approach can be regarded as the more exact method for describing the energy spectra of considered nuclei in transitional region.

### 3.2. *B*(*E*2)Transition probabilities

The stable even-even nuclei in *Ba* isotopic chain exhibit an excellent opportunity for studying behavior of the total low-lying *E*2 strengths in transitional region from deformed to spherical nuclei. Computation of electromagnetic transition is a sign of good test for nuclear model wave functions. To determine boson effective charges, we have used same method introduced in Refs.[11-12,34-35]. We have extracted these quantities from the empirical *B*(*E*2) values via Least square technique. The parameters of Eqs. 43, effective charges, have been presented in Tables 3 and 4 for IBM-1 and IBM-2 predictions, respectively.

In IBM-2 formalism, we have used a similar method which finally the following values are yield via extraction procedures for quadrupole coefficients.

We have presented the predictions of IBM-1 and IBM-2 for some of quadrupole transitions which their experimental counterparts are available presented in Table5. Our results suggest better agreement between experimental data with predictions of IBM-2 in comparison with IBM-1. Also, for interband transitions, theoretical predictions are more exact in comparison with intraband ones.

From these Figures and Tables, one can conclude, the calculated energy spectra in this approach are generally in good agreement with the experimental data. Our results indicate the elegance of the extraction procedure which has been presented in this technique and they suggest the success of estimation processes. Also, theoretical *B*(*E*2) transition probabilities of even-even *Ba* isotopes, which have been obtained by using the model perspectives, exhibit nice agreement with experimental ones. We would extend this study to consider other nucleus in this isotopic chain which the concept of shape coexistence and critical behavior in this transitional region are important for them.

## 4. CONCLUSIONS

In this paper, we have studied the $^{124-130}Ba$ isotopic chain in the $U(5) \leftrightarrow SO(6)$ transitional region of interacting boson model via a $SU(1,1)$-based transitional Hamiltonian. We have used the catastrophe theory in combination with coherent state formalism to generate energy surfaces and determine the exact values of control parameters. Energy levels and $B(E2)$ transition probabilities are determined in the both IBM-1 and 2 formalism and it is seen that there is an existence of a satisfactory agreement between presented results and experimental counterparts. We may conclude that general characteristics of *Ba* isotopes are well accounted in this study and the idea for combination of spherical and gamma unstable shapes is supported in this region. The obtained results in this study confirm that this technique is worth extending for investigating the nuclear structure of other nuclei existing around the mass of $A \sim 130$.

### Acknowledgement

One of the authors (H.S) is grateful to H. fathi for his assist in the determination of energy surfaces. This work is published as a part of research project supported by the University of Tabriz Research Affairs Office.

Tables

Table1. Parameters of IBM-1 Hamiltonian are showed for different *Ba* isotopes. *N* describes the boson number and σ regards as the quality of extraction processes.

| Nucleus | $N$ | $\varepsilon(kev)$ | $c_s$ | $\gamma(kev)$ | $\delta(kev)$ | $\sigma$ |
|---|---|---|---|---|---|---|
| $^{124}_{56}Ba$ | 10 | 284.8 | 0.48 | -10.01 | 12.19 | 197.5 |
| $^{126}_{56}Ba$ | 9 | 251.3 | 0.60 | 21.97 | -3.37 | 171.6 |
| $^{128}_{56}Ba$ | 8 | 268.8 | 0.41 | 21.98 | 3.68 | 185.1 |
| $^{130}_{56}Ba$ | 7 | 391.7 | 0.38 | -17.82 | 3.58 | 195.2 |

Table2. Parameters of IBM-2 Hamiltonian are presented for different *Ba* isotopes. *N* describes boson number and σ regards as the quality of extraction processes.

| Nucleus | $N$ | $\varepsilon(kev)$ | $c_s$ | $\gamma_1(kev)$ | $\gamma_2(kev)$ | $\delta_1(kev)$ | $\delta_2(kev)$ | $\delta(kev)$ | $\sigma$ |
|---|---|---|---|---|---|---|---|---|---|
| $^{124}_{56}Ba$ | 10 | 177.2 | 0.55 | 40.0 | 16.0 | -51.7 | -5.1 | 5.6 | 79.1 |
| $^{126}_{56}Ba$ | 9 | 405.3 | 0.92 | -0.4 | -211.1 | -201 | -254 | 202.8 | 18.7 |
| $^{128}_{56}Ba$ | 8 | 239.5 | 0.81 | 34.8 | -111.9 | -12.5 | -55.1 | 10.6 | 88.6 |
| $^{130}_{56}Ba$ | 7 | 244.3 | 0.41 | -3.4 | 26.1 | -0.5 | -19.2 | 0.75 | 8.8 |

Table3. Predictions of IBM-1 presented for effective quadrupole coefficients of considered isotope. $^{124}Ba$ doesn't have enough experimental data for analysis.

| Nucleus | $^{126}_{56}Ba$ | $^{128}_{56}Ba$ | $^{130}_{56}Ba$ |
|---|---|---|---|
| $q_2$ | 7.4010 | 0.3950 | $1.5 \times 10^{-8}$ |
| $q_2'$ | $-6.78 \times 10^{-9}$ | $9.0 \times 10^{-9}$ | 1.6927 |

Table4. IBM-2 predictions are presented for effective quadrupole coefficients. $^{124}Ba$ doesn't have enough experimental data for analysis.

| Nucleus | $^{126}_{56}Ba$ | $^{128}_{56}Ba$ | $^{130}_{56}Ba$ |
|---|---|---|---|
| $e_\pi$ | 1.0432 | 0.8066 | 1.5836 |
| $e_\nu$ | 0.0877 | 1.0331 | 0.3411 |
| $q_2$ | 2.3980 | 1.9076 | 2.4462 |
| $q_2'$ | -0.2437 | -2.1565 | -2.2695 |

Table5. The predictions of IBM-1 and 2 are presented for some quadrupole transition probabilities which their experimental counterparts are available from Refs.[42-46]. All quantities are expressed in W.u.

| Nucleus | transition | Experimental | IBM-1 | IBM-2 |
|---|---|---|---|---|
| $^{126}_{56}Ba$ | $2^+_1 \to 0^+_1$ | 94 | 123 | 105 |
| | $4^+_1 \to 2^+_1$ | 128 | 145 | 136 |
| | $6^+_1 \to 4^+_1$ | 173 | 190 | 178 |
| | $2^+_1 \to 0^+_1$ | 200 | 214 | 206 |
| $^{128}_{56}Ba$ | $2^+_1 \to 0^+_1$ | 72 | 84 | 77 |
| | $4^+_1 \to 0^+_1$ | 18 | 23 | 21 |
| | $2^+_2 \to 0^+_1$ | 3.4 | 3.99 | 3.75 |
| | $4^+_2 \to 2^+_2$ | 62 | 71 | 68 |
| | $4^+_2 \to 2^+_1$ | 0.9 | 1.45 | 1.22 |
| | $6^+_1 \to 4^+_1$ | 101 | 111 | 107 |
| | $8^+_1 \to 6^+_1$ | 96 | 104 | 100 |
| | $6^+_2 \to 4^+_2$ | 100 | 108 | 102 |
| | $6^+_2 \to 2^+_2$ | 0.78 | 0.98 | 0.86 |
| $^{128}_{56}Ba$ | $2^+_1 \to 0^+_1$ | 57.9 | 68.1 | 60.5 |
| | $4^+_1 \to 2^+_1$ | 78.9 | 86.3 | 81.4 |
| | $6^+_1 \to 4^+_1$ | 94 | 106 | 98 |
| | $8^+_1 \to 6^+_1$ | 90 | 97 | 92 |
| | $10^+_1 \to 8^+_1$ | 54 | 67 | 58 |
| | $12^+_1 \to 10^+_1$ | 24 | 33 | 28 |

# Figure caption

Figure1. IBM-1 predictions for energy levels of $^{124-130}Ba$ isotopes and their experimental counterparts were taken from Refs.[27-30].

Figure2. IBM-2 predictions for energy levels of $^{124-130}Ba$ isotopes and their experimental counterparts were taken from Refs.[27-30].

Figure3. Energy surfaces for *Ba* isotopes which are derived by the predictions of IBM-2. The $^{126}Ba$ has the most similarity with the expected figure for *SO*(6) limit.

Figure1.

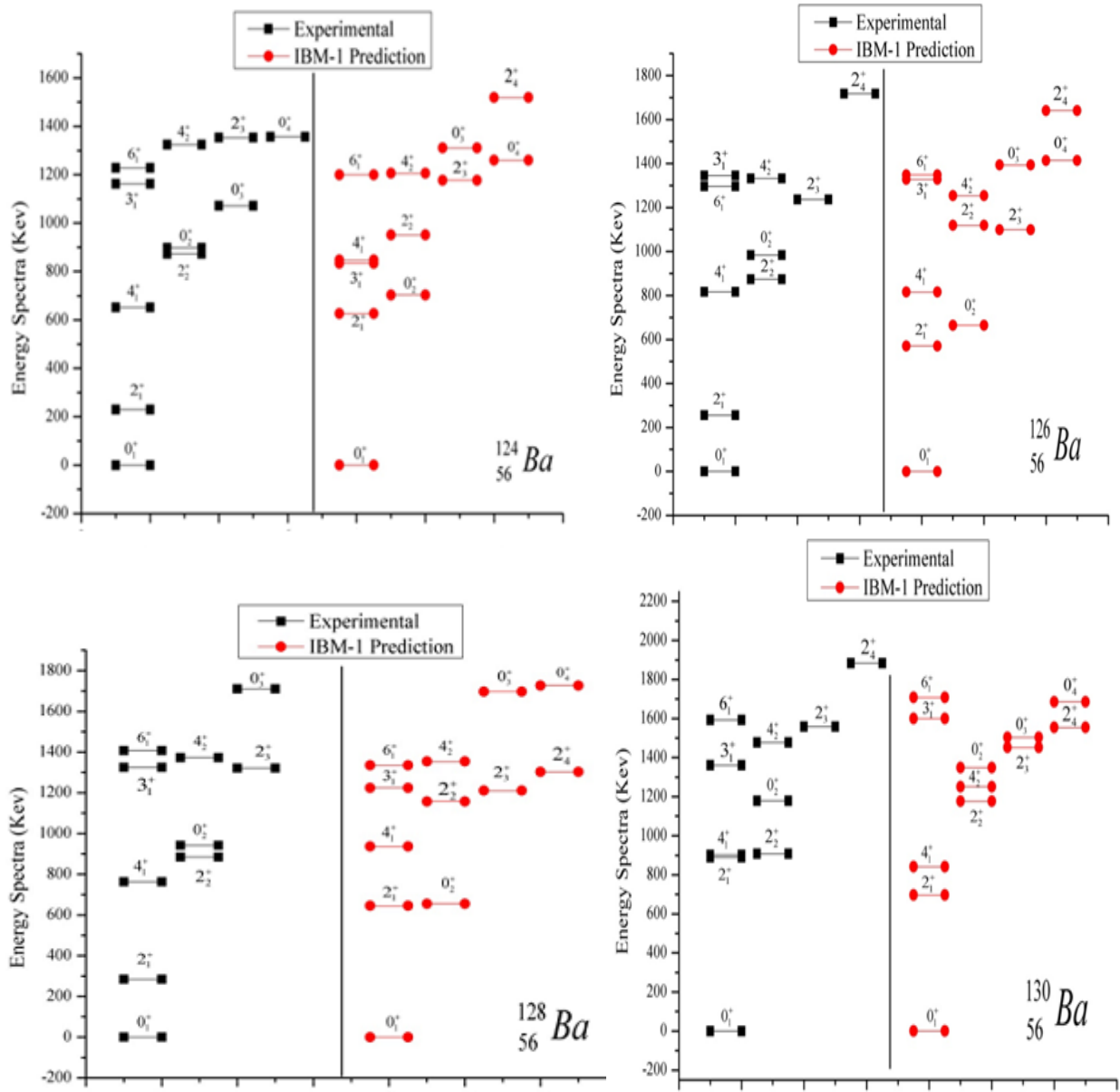

Figure2.

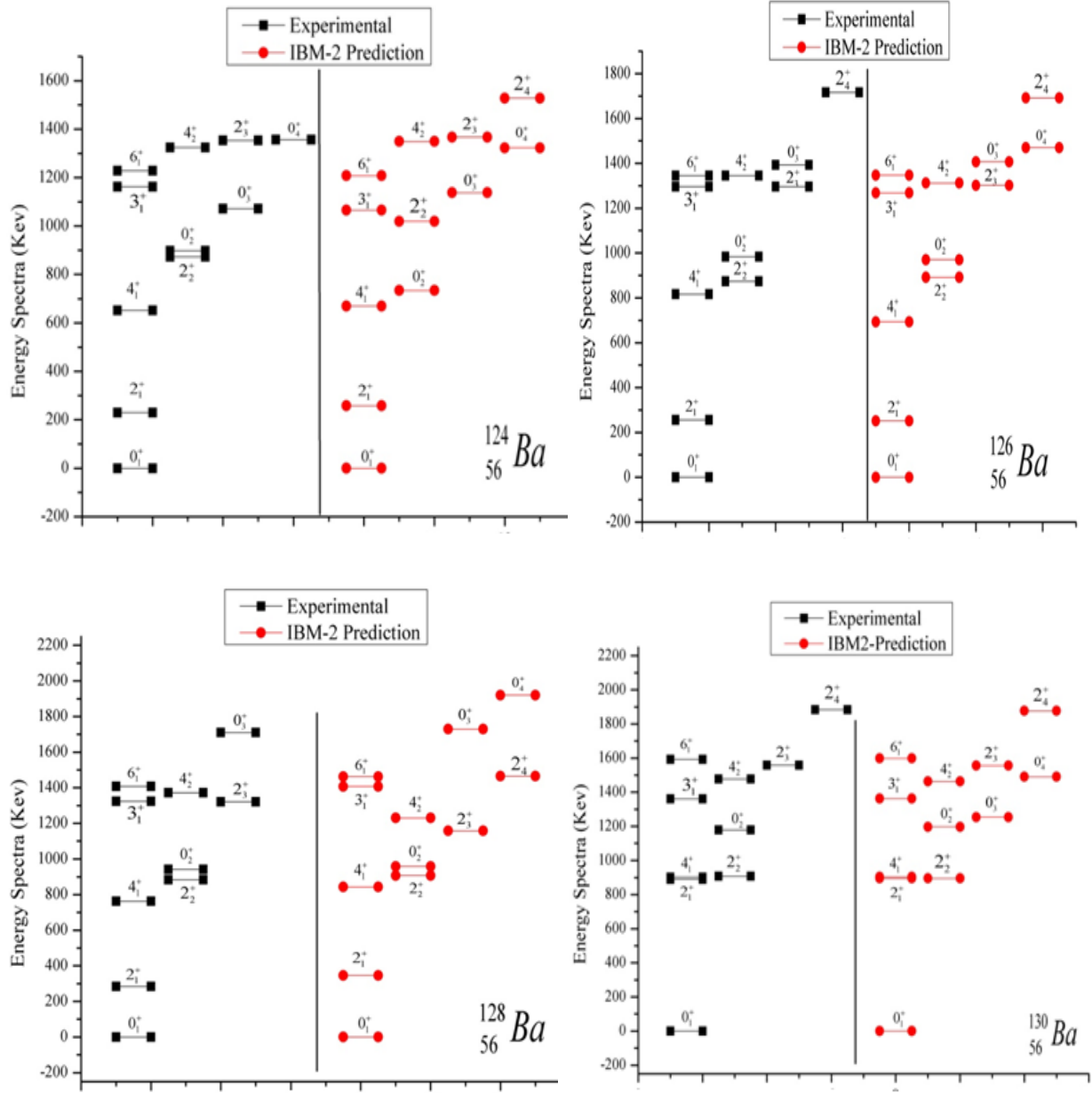

Figure3.

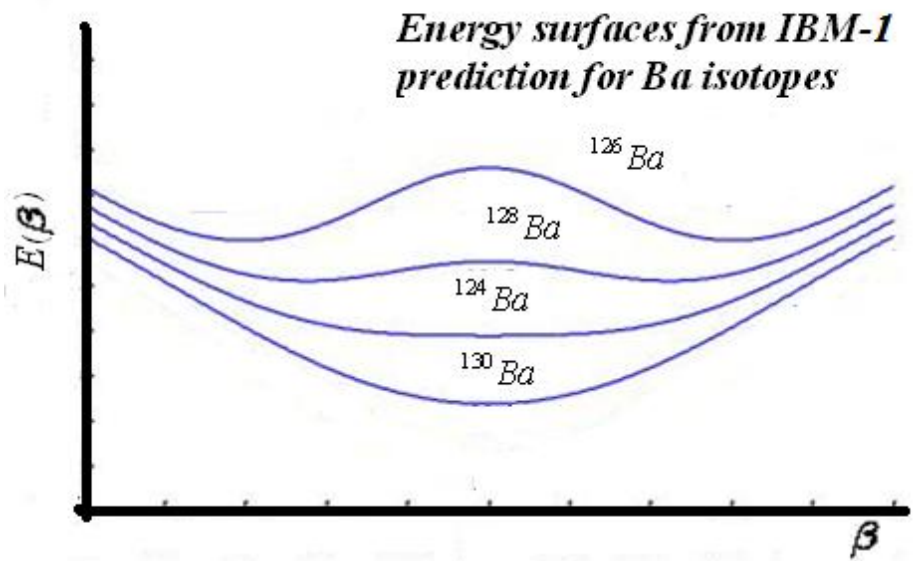